\documentclass{aa}
\usepackage[dvips]{graphicx}
\usepackage{psfig}
\usepackage{ltxtable}
\def\lesssim{\mathrel{\hbox{\rlap{\hbox{\lower4pt\hbox{$\sim$}}}\hbox{$<$}}}}
\def\gtrsim{\mathrel{\hbox{\rlap{\hbox{\lower4pt\hbox{$\sim$}}}\hbox{$>$}}}}
\begin{document}
\title{Freeze--out and coagulation in pre--protostellar collapse}


\author{D.R. Flower\inst{1}
\and G. Pineau des For\^{e}ts\inst{2,3}
\and C.M. Walmsley\inst{4}}

\institute{Physics Department, The University,
           Durham DH1 3LE, UK
\and       IAS, Universit\'{e} de Paris--Sud, F-92405 Orsay, France
\and       LUTH, Observatoire de Paris, F-92195, Meudon Cedex, France
\and       INAF, Osservatorio Astrofisico di Arcetri,
           Largo Enrico Fermi 5, I-50125 Firenze, Italy}

\offprints{C.M. Walmsley}


   \abstract{ We study the changes in physical and chemical conditions
during the early stages of collapse of a pre--protostellar core, starting from initial conditions appropriate to a dense molecular cloud and proceeding to the ``completely depleted'' limit. We allow for molecular desorption from the grain surfaces and follow the evolution of the ionization degree and the ionic composition as functions of time and density. The timescale for collapse is treated as a parameter and taken equal to 
either the free--fall or the ambipolar diffusion time. The processes of freeze--out on to the dust grains and of coagulation of the grains were treated simultaneously with the chemical evolution of the medium in the course of its collapse. When proceeding at close to its maximum rate, coagulation has important consequences for the degree of ionization and the ionic composition of the medium, but its effect on the freeze--out of the neutral species is modest. An innovation of our study is to calculate the grain charge distribution; this is done in parallel with the chemistry and the dynamics. The grain charge distribution is significant because H$^+$ ions recombine predominantly on the surfaces of negatively charged grains. We have also attempted to reproduce with our models the observational result that
nitrogen--containing species, such as NH$_3$ and N$_2$H$^+$, remain in the gas phase at densities for which CO and other C--containing molecules appear to have frozen on to grain surfaces.  We conclude that recent measurements of the adsorption energies of N$_2$ and CO invalidate the interpretation of these observations in terms of
the relative volatilities of N$_2$ and CO. We consider an alternative explanation, 
in terms of low sticking coefficients for either molecular or atomic N; but this hypothesis requires experimental confirmation. We find that, irrespective of the nitrogen chemistry, the main gas phase ion is
either H$^+$ or H$_3^+$ (and its deuterated isotopes) at densities above $10^5$ cm$^{-3}$; whether H$^+$ or H$_3^+$  predominates depends sensitively on the rate of increase in grain size (decrease
in grain surface area per unit volume of gas) during core contraction. Our calculations show that H$^+$  will predominate if grain coagulation proceeds at close to its maximum rate, and H$_3^+$ otherwise. 

            }

\maketitle
%

\section{Introduction}

The contraction of an interstellar ``core'' to form a
protostar is a process of great current interest. It is clear 
that this contraction is accompanied by considerable changes
both in the characteristics of the grains associated with the
core and the chemical composition of the gas. In particular, 
most molecules containing heavy elements condense
on to grain surfaces (Bacmann et al.~2002, Tafalla et al.~2004). 
For CO and several other species, this process occurs at densities $n_{\rm H} 
\gtrsim 3\, 10^4$ cm$^{-3}$. Moreover, there is 
evidence in several objects for a ``molecular hole'' 
at densities above $10^6$ cm$^{-3}$, where 
material containing C,N,O is in  solid form
and only molecules and ions composed of H and D remain in the
gas phase.  

In two earlier articles (Walmsley et al.~2004, Flower et al.~2004), we have considered the characteristics of such ``completely depleted''
regions and have concluded
that the degree of ionization of the gas and the relative
abundances of the major ions (H$^+$, H$_3^+$, H$_2$D$^+$, D$_2$H$^+$, D$_3^+$) are sensitive to the size of the negatively charged grains, with which H$^+$ ions recombine.
Recent observations  of H$_2$D$^+$ and D$_2$H$^+$
(Caselli et al.~2003, Vastel et al.~2004)
are compatible with the existence of a zone lacking heavy
elements. However, detection of these ions (H$_2$D$^+$ and D$_2$H$^+$) 
seems to be rare, suggesting that 
objects containing such a depleted nucleus are in the ultimate
phase of evolution, prior to the formation of a protostar. 
It seems relevant also that the observations suggest
that there is an intermediate density range ($10^5 \lesssim n_{\rm H} 
\lesssim 10^6$ cm$^{-3}$) where 
molecules containing nitrogen, such as NH$_3$ and
N$_2$H$^+$, remain in the gas phase when carbon-- and oxygen--containing species are
already depleted out. This effect of differential depletion was believed to be due to the relatively high volatility
of N$_2$ (see Bergin \& Langer 1997), which remains 
in the gas phase when CO, for example, has condensed
out. However, recent measurements of the
adsorption energies of CO and N$_2$, which are found to differ by less than 10\% (Oberg et al. 2004), do not support this interpretation. Accordingly, we attempt to account for the differential depletion by varying other (uncertain) parameters of the model. 

An important issue is the extent to which grain coagulation 
takes place during
contraction, thereby influencing the timescale 
for molecular depletion.  Studies of this process (e.g. Ossenkopf
1993, Ossenkopf \& Henning 1994, Suttner \& Yorke 2001)
suggest that, at densities $10^5 \lesssim n_{\rm H} 
\lesssim 10^7$ cm$^{-3}$, grains grow in size, 
although not to the same extent as in protoplanetary disks, where the density is higher.  The grain opacity per H nucleus, defined here as $(n_{\rm g}/n_{\rm H})\pi a_{\rm g}^2$, where $n_{\rm g}$ is the number density of grains of radius $a_{\rm g}$ and $n_{\rm H} = n({\rm H}) + 2n({\rm H}_2)$, is expected to decrease but not sufficiently to prevent depletion on to grain surfaces (whose rate is proportional to  the same parameter); this is consistent with what is
known about the mm--sub-mm opacity of prestellar cores
(e.g. Bianchi et al.~2003, Kramer et al.~2003). 

With the above points in mind, we have undertaken calculations 
of the evolution of the chemical composition of such a contracting core. 
Our work is similar in some respects, but differs in others, to studies of the surface and gas--phase chemistries in a contracting core by Aikawa et al. (2001, 2003, 2005), as well as the time--dependent calculations of Roberts et al. (2003) and the recent study of Bonnor--Ebert spheres by Lee et al. (2004). Our model incorporates a less detailed treatment of the gas--phase chemistry, and we do not attempt to treat surface reactions explicitly. Instead, we have emphasized the evolution of grain charge and size with time and the consequences for the ion chemistry, which we perceive to be a key aspect of the problem. The grain size (through the grain surface area) influences the relative abundances of H$^+$ and H$_3^+$, which are likely to be the major ions during the later phases of core contraction. Our models bring out the important role played under some circumstances by H$^+$, whose significance appears to have been neglected -- or, at least, not emphasized -- in earlier studies. 

In Section 2 of this article, we summarize our formulation of
the problem. In Section 3, we consider the critical input
parameters of the model. Our main results are given in Section 4 and our 
concluding remarks in Section 5. 

\section{Formulation}

We consider the evolution of a contracting sphere (``core'') of gas and dust and, 
for simplicity, we use equations applicable
to free--fall, homologous, isothermal collapse.  This  procedure 
allows us
to parametrize the dynamical behaviour in terms of a collapse
timescale, $\tau _{\rm c}$, which we vary between a minimum value,
taken to be the  
free--fall time, $\tau _{\rm ff}$, and a maximum value, defined
by the ambipolar diffusion timescale, $\tau _{\rm ad}$ (see
below), both evaluated according to the initial conditions in the medium. In practice, the ambipolar timescale is about an order
of magnitude larger than the free--fall time. Thus,
$\tau _{\rm c}$ spans timescales between the magnetically
subcritical case, in which the magnetic flux suffices to prevent
collapse, and the magnetically supercritical case, where collapse
can take place immediately.

The equation which determines the variation of the radius $R$ of the core with time $t$ is

\begin{equation}
\frac {1}{R} \frac {{\rm d}R}{{\rm d}t} \equiv \frac {1}{x} \frac {{\rm d}x}{{\rm d}t} = -\frac {\pi }{2 \tau _{\rm c} x} \left (\frac {1}{x} - 1 \right )^{\frac {1}{2}}
\label{equ1}
\end{equation}
where $x = R/R_0 \le 1$ and $R_0$ is the initial radius of the condensation. The timescale for free--fall collapse is (Spitzer 1978)

\begin{equation}
\tau _{\rm ff} = \left [ \frac {3\pi }{32G\rho _0} \right ]^{\frac {1}{2}}
\label{equ2}
\end{equation}
where $\rho _0$ is the initial mass per unit volume. In spherical collapse, the mass density $\rho $ varies with time $t$ according to

\begin{equation}
\frac {1}{\rho } \frac {{\rm d}\rho }{{\rm d}t} = - \frac {3}{x} \frac {{\rm d}x}{{\rm d}t} 
\label{equ3}
\end{equation}
In practice, the collapse is described by equation (\ref{equ1}), with the timescale $\tau _{\rm c}$ taken to be either $\tau _{\rm ff}$ or the ambipolar diffusion timescale, $\tau _{\rm ad}$, to be defined below. As already noted, $\tau _{\rm ad}$ is about an order of magnitude larger than $\tau _{\rm ff}$.

The medium contains neutral gas, positive ions and electrons; it also contains neutral and (mainly negatively) charged grains. The number density, $n$, of the gas--phase neutrals is determined by

\begin{equation}
\frac {1}{n} \frac {{\rm d}n}{{\rm d}t} = - \frac {3}{x} \frac {{\rm d}x}{{\rm d}t} + \frac {\cal N}{n}
\label{equ6}
\end{equation}
where ${\cal N}$ is the rate per unit volume at which neutrals are created in the gas phase by chemical reactions. Similarly, for the positive ions in the gas phase

\begin{equation}
\frac {1}{n_+} \frac {{\rm d}n_+}{{\rm d}t} = - \frac {3}{x} \frac {{\rm d}x}{{\rm d}t} + \frac {\cal N_+}{n_+}
\label{equ7}
\end{equation}
Analogous equations are solved for each of the chemical species which intervene in the chemistry.

Central to our approach, and distinguishing it from previous work, is the self--consistent calculation of the change in the grain cross section which occurs owing to coagulation and to freeze--out from the gas phase. 
The grains are assumed to have a unique radius, determined initially by the grain:gas  mass ratio and the mass density of the grain material. Subsequently, the grain radius varies owing to the freeze--out of gas--phase constituents on to the surfaces of the grains and to grain coagulation. The rate of variation of the grain number density is given by

\begin{equation}
\frac {1}{n_{\rm g}} \frac {{\rm d}n_{\rm g}}{{\rm d}t} = - \frac {3}{x} \frac {{\rm d}x}{{\rm d}t} + \frac {\cal N_{\rm g}}{n_{\rm g}}
\label{equ8}
\end{equation}
where $n_{\rm g}$ is the grain number density. The first term on the right hand side of this equation represents the rate of increase of the number density owing to the compression associated with the (spherical) collapse; the second represents the rate of decrease of the grain number density owing to coagulation. When two grains stick together, only one grain remains, and hence the rate per unit volume at which the grains are removed through coagulation is ${\cal N}_{\rm g} = - \langle \sigma v \rangle n_{\rm g}^2$. The method used to determine the rate constant $\langle \sigma v \rangle$ is presented in Appendix A. We have assumed the grains to be spherical, which minimizes their surface area and effective cross section for a given volume. Clearly, non--spherical grains would have a larger cross section.  

In addition to the free--fall timescale [equation (\ref{equ2})], we evaluate the ambipolar diffusion timescale from

\begin{equation}
\tau _{\rm ad} =  \frac {2}{\pi Gm_{\rm n}^2} \sum _{\rm i} \frac {n_{\rm i}}{n_{\rm n}} \frac {m_{\rm i}m_{\rm n}}{m_{\rm i}+m_{\rm n}}                \langle \sigma v \rangle _{\rm in}
\label{equ9}
\end{equation}
(cf. Flower \& Pineau des For\^{e}ts 2003a) for both positive ions and charged grains (in which case $m_{\rm i} >> m_{\rm n}$) and adopt the larger of the two values. We use the expression of Osterbrock (1961) for the rate coefficient for momentum transfer in a collision between an ion and a neutral particle,

\begin{equation}
\langle \sigma v \rangle _{\rm in} = 2.41\pi (e^2\alpha _{\rm n}/m_{\rm in})^{\frac{1}{2}}
\label{equ10}
\end{equation}
where $m_{\rm in} = m_{\rm i}m_{\rm n}/(m_{\rm i} + m_{\rm n})$ is the reduced mass and $\alpha _{\rm n}$ is the polarizability of the neutral. For collisions between neutrals and charged grains, we take $\sigma = \pi a_{\rm g}^2$ and $v = (8k_{\rm B}T_{\rm n}/\pi m_{\rm n})^{\frac{1}{2}}$, where $T_{\rm n}$ is the temperature of the neutral gas. 

\section{Model}

Adopting the initial conditions in steady--state, the condensation was allowed to undergo isothermal collapse, as described in Section 2. The dynamical and chemical rate equations were solved simultaneously. The grain number density was determined by the competing effects of compression, arising from the collapse, and coagulation, which tends to reduce the total number of grains [cf. equation~(\ref{equ8}) above]. The grain radius was calculated self--consistently as a function of time, $t$, from the grain number density and the total mass of material currently in the solid phase, allowing for the continuing adsorption of heavy species on to the grains. In practice, adsorption has only a minor effect on $a_{\rm g}$, as a mantle of ices is assumed to exist before the onset of collapse; coagulation is a much more significant mechanism for increasing $a_{\rm g}$. 

\subsection{Initial conditions}

The initial distribution of the elemental abundances across the gas phase and the solid phase (i.e. grains) is given in Table~\ref{elements}. The elemental abundances derive from the study of Anders \& Grevesse (1989), and the depletions from Savage \& Sembach (1996), Gibb et al. (2000), and Sofia \& Meyer (2001), with the exception of sulphur, for which the depletion depletion from the gas phase within molecular clouds is difficult to establish observationally. We have assumed that sulphur is depleted from the gas phase by a factor of 30 (cf. Wakelam et al. 2004). Given the distribution in Table~\ref{elements}, the initial value of the mass of dust to the mass of gas is 0.0094.

The computer code was run to determine the steady--state abundances of the chemical species in the gas phase (neglecting any further depletion) by the simple expedient of specifying a timescale for collapse, $\tau _{\rm c}$, in equation~(\ref{equ1}) which is much greater than the chemical equilibrium timescale; 92 gas--phase species and almost 800 chemical reactions were included in the model. The gas--phase chemistry included ion--neutral, neutral--neutral, and dissociative recombination reactions involving species containing H, He, C, N, O and S. The branching ratio for the dissociative recombination of N$_2$H$^+$, to N or N$_2$, was taken from the recent measurements of Geppert et al. (2004). Ionization by cosmic rays, both direct and indirect (by the electrons released), was taken into account. Adsorption of neutral species on to grains, and desorption of the hydrogenated products of the grain--surface chemistry following cosmic ray heating of the grains, were also incorporated (see below). The rate of cosmic ray ionization of hydrogen was taken to be $\zeta = 1\times 10^{-17}$ s$^{-1}$. The list of species and the complete set of chemical reactions included in the model are available from http://massey.dur.ac.uk/drf/protostellar/species\_chemistry.

\begin{table}
\caption{ The elemental abundances and the initial repartition between the gas phase and the solid phase: based on the studies of Anders \& Grevesse (1989), Savage \& Sembach (1996), Gibb et al. (2000), and Sofia \& Meyer (2001). `Depletion' denotes the fraction of the element in the solid phase (grains). Numbers in parentheses are powers of 10. }
\vspace{1em}
\begin{tabular}{lllll}
\hline
  Element & Fractional abundance & Gas phase & Solid phase & Depletion  \\
\hline
\hline
H &  1.00 &  1.00 & 3.95(-4) & 0.0  \\
He &  1.00(-1) &  1.00(-1) & & 0.0  \\
C &  3.55(-4) &  8.27(-5) &  2.72(-4) & 0.77  \\
N &  7.94(-5) &  6.39(-5) & 1.55(-5) & 0.20   \\
O &  4.42(-4) &  1.24(-4) &  3.18(-4) & 0.72  \\
Mg &  3.70(-5) &   &  3.70(-5) & 1.0  \\
Si &  3.37(-5) &   &  3.37(-5) & 1.0  \\
S &  1.86(-5) &  0.60(-6) &  1.80(-5) & 0.97  \\
Fe &  3.23(-5) &  1.50(-9) & 3.23(-5) & 1.0  \\
\hline
\end{tabular}
\label{elements}
\end{table}

The initial value of the grain radius was taken to be $a_{\rm g} = 0.05$ $\mu $m, which yields a grain opacity, defined here as $n_{\rm g}\pi a_{\rm g}^2$, of  $1.6\times 10^{-21}n_{\rm H}$ cm$^{-1}$, assuming the density of the grain material to be 2 g cm$^{-3}$. Adopting the grain--size distribution of Mathis et al. (1977), with limits of $0.01 \le a_{\rm g} \le 0.3$ $\mu $m to the distribution, and assuming the canonical value of 0.01 for the dust:gas mass ratio, one obtains the same value of $n_{\rm g}\pi a_{\rm g}^2$. [We note that the value of $a_{\rm g}$ required to reproduce a given value of $n_{\rm g}\pi a_{\rm g}^2$ depends only on the adopted size distribution and its limits, and not on the density of the grain material nor the dust:gas mass ratio: see Appendix B.]

\subsection{Grain charge distribution}

In view of the significance of grains for the state of charge of the medium, the grain charge distribution was computed in parallel with the chemical composition, allowing for electron attachment and detachment and electron--ion recombination on grain surfaces (cf. Flower \& Pineau des For\^{e}ts 2003b, Section 2.2). Singly charged and neutral grains were considered. In the case of collisions betwen charged particles and oppositely charged grains, the `focusing' effect of the Coulomb attraction was included by means of equation~(3.4) of Draine \& Sutin (1987). In the case of collisions between charged particles and neutral grains, the attraction arising from the polarization of the grain by the incoming particle was included by means of their equation~(3.3). For example, the rate per unit volume of electron attachment to neutral grains was evaluated as $n_{\rm e}n_{{\rm g}^0}\pi a_{\rm g}^2[1 + (\pi e^2/2a_{\rm g}k_{\rm B}T_{\rm e})^{0.5}][8k_{\rm B}T_{\rm e}/(\pi m_{\rm e})]^{0.5}s_{\rm e}$, where $n_{\rm e}$ is the electron density, $n_{{\rm g}^0}$ is the density of the neutral grains, $T_{\rm e}$ is the electron temperature (taken equal to the gas temperature, $T = 10$~K), $m_{\rm e}$ is the electron mass, and $s_{\rm e} = 0.5$ is the electron sticking coefficient. The initial value of the total grain number density, $n_{\rm g}$,  was calculated self--consistently from the elemental depletions, the specified grain radius, and the mean density of the grain material, taken to be 2 g cm$^{-3}$ at all times. The grain charge distribution is not sensitive to the assumed grain size, as the dominant charge--changing processes -- electron attachment to neutral grains and ion recombination with negatively charged grains -- scale approximately with $a_{\rm g}^{\frac{3}{2}}$ and with $a_{\rm g}$, respectively, under the conditions which are relevant here.

\subsection{Adsorption and desorption}

The rate coefficient for adsorption of neutral species on to the grains was taken to be $\pi a_{\rm g}^2[8k_{\rm B}T/(\pi m_{\rm n})]^{0.5}S_{\rm n}$, where $m_{\rm n}$ is the mass of the neutral. The `sticking' coefficient $S_{\rm n} = 1$, unless stated otherwise; we assumed that He remained in the gas phase and that H$_2$ was returned to the gas phase after forming on the grains. The geometrical cross section of the grain is adopted in this expression for the rate coefficient. Although the potential arising from the polarization of the grain by a charged particle can be significant (see Section 3.2), the potential due to the polarization of a neutral by a charged grain is negligible; this is because the polarizability of a grain is orders of magnitude larger than the polarizabilities of the neutral gas--phase species, owing to the much larger size of the grain.

The recombination of positive ions with electrons on the surfaces of negatively charged grains releases the electron binding energy, which is typically a few eV and large compared with the adsorption energies of the neutral species. Accordingly, we have assumed that the neutrals which are formed in this way are returned immediately to the gas phase. The attachment of ions to neutral grains is a relatively minor process, owing to the predominance of negatively charged grains (see Section 4.3 below) and to Coulomb focusing in collisions of ions with negatively charged grains.  Consequently, we have neglected the adsorption of ions on to grains when computing the rate of freeze--out of gas--phase species.

With the exception of H$_2$, which desorbs following its formation, desorption of molecules was assumed to be controlled by the heating of grains which occurs owing to cosmic ray impact (L\'{e}ger et al. 1985). The rate of desorption of species $i$ per unit volume per unit time from the grains is

\begin{displaymath}
\frac {n_{\rm i}^{g}}{\sum _{\rm i} n_{\rm i}^{g}} n_{\rm g} 4\pi a_{\rm g}^2\ \gamma \ {\rm exp} [\frac {-(E_{\rm i}^{\rm ads} - E_{{\rm CO}}^{\rm ads})}{T_{\rm g}^{\rm max}}]
\end{displaymath}
where $n_{\rm i}^{g}/\sum _{\rm i} n_{\rm i}^{g}$ is the fractional abundance of species $i$ on grains, $E_{\rm i}^{\rm ads}$ is the adsorption energy of the species (Aikawa et al. 1996, 1997, 2001), which relate to measurements on pure ices; $T_{\rm g}^{\rm max} = 70$~K is the maximum temperature reached by the grains following cosmic ray impact (Hasegawa \& Herbst 1993).  We adopted $\gamma = 70$ cm$^{-2}$ s$^{-1}$, which yields a rate of desorption for CO which is consistent with the value for `spot' heating, as derived by L\'{e}ger et al. (1985); the heating arises from the impact of heavy cosmic rays, particularly Fe nuclei. In the case of small grains,  `whole--grain' heating may be more important than `spot' heating, and the corresponding value of $\gamma $ would be larger than adopted here.

\subsection{Coagulation}

Grain coagulation was treated as described in Appendix A.
The important parameters are the local turbulent
speed of the gas, which is assumed to determine the relative
speed of the grains, and the critical speed $v_{\rm crit}(a_{g})$
below which grains are assumed to coagulate on every collision.  Following     Tafalla et al. (2004), who observed L1498 and L1517B (see also Barranco \& Goodman 1998), we adopted the turbulent line--width as an upper limit to the FWHM of the Gaussian velocity distribution, i.e. $\Delta v \lesssim 0.12$ km s$^{-1}$, which is approximately $c_{\rm s}/2$ at $T = 10$~K, where $c_{\rm s}$ is the adiabatic sound speed.  This value may include contributions from 
motions on scales larger than the mean--free path for
grain--grain collisions and hence places an upper limit on $\Delta v$.  Accordingly, we have taken $\Delta v = c_{\rm s}/4$ as our reference value, but we have also run models with $\Delta v = c_{\rm s}/2$ and $\Delta v = c_{\rm s}/10$.

The critical speed for coagulation, $v_{\rm crit}$, was taken from Chokshi et al. (1993), unless otherwise stated. We adopted their results for the case of `ice'; this seems the most appropriate of the three cases (`ice', `graphite', `quartz') considered by Chokshi et al., as the grains already have ice mantles at the start of the calculation. Accordingly,

\begin{equation}
v_{\rm crit}(a_{\rm g}) = 0.1\ a_{\rm g}^{-5/6}
\label{equ11}
\end{equation}
when $v_{\rm crit}$ is in cm s$^{-1}$ and $a_{\rm g}$ is in cm. We have assumed that the grains remain spherical throughout. This assumption minimizes the grain radius, and hence the grain opacity per H nucleus, $(n_{\rm g}/n_{\rm H})\pi a_{\rm g}^2$, as the collapse proceeds. We assume also that the density of the grain material is unchanged by coagulation. The critical speed given by equation~(\ref{equ11}) is greater than the estimate of Suttner \& Yorke (2001) but approximately 4 times less than the measurement of Poppe \& Blum (1997). We have run some comparison calculations in which $v_{\rm crit}$ was taken to be 4 times greater than is given by equation~(\ref{equ11}).

\section{Results}

We consider the case of a condensation of gas and dust with an initial number density $n_{\rm H} = n({\rm H}) + 2n({\rm H}_2) = 10^4$ cm$^{-3}$ and kinetic temperature $T = 10$~K. We recall that the rate of cosmic ray ionization of hydrogen was taken to be $\zeta = 1\times 10^{-17}$ s$^{-1}$.

\subsection{Freeze--out}

\begin{figure}
\centering
\includegraphics[height=20cm]{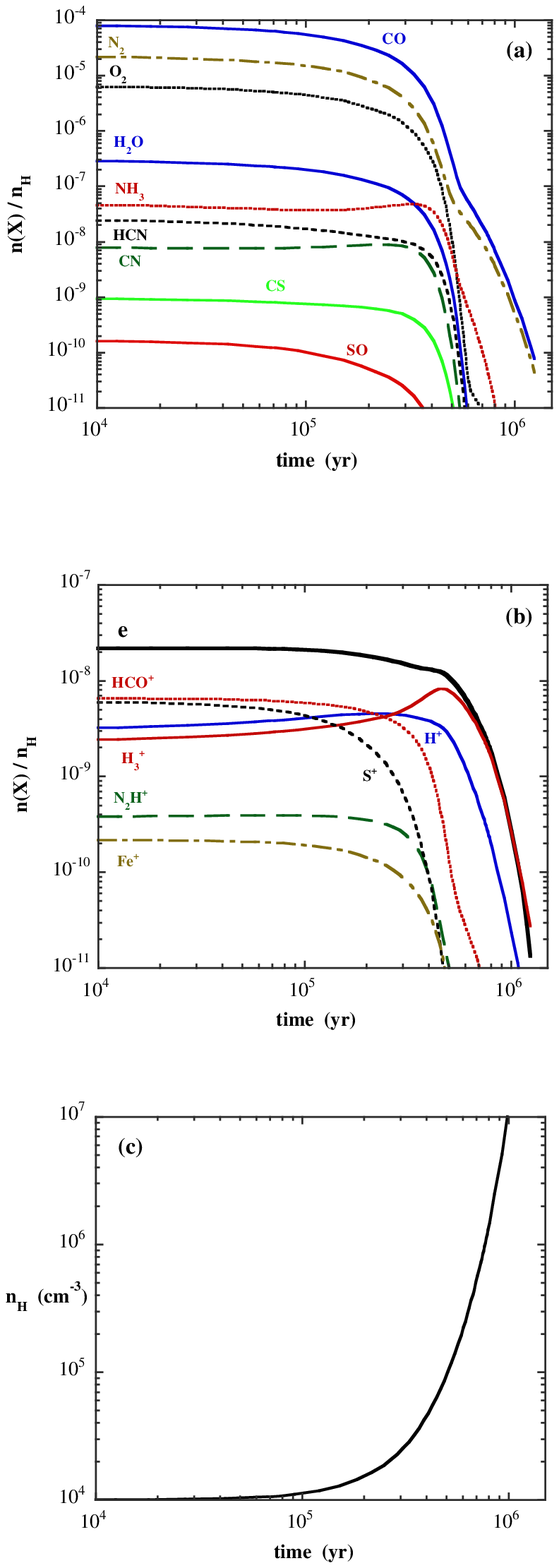}
\caption{The fractional abundances of (a) selected neutrals and (b) the major ions, as functions of the evolution time, $t$. Results are shown for the limiting case of no coagulation and assuming that the timescale for collapse is determined by the free--fall time, $\tau _{\rm ff}$. The corresponding density profile, $n_{\rm H}(t)$, is plotted in panel (c).}
\label{f-o}
\end{figure}

Fig.~\ref{f-o} illustrates the freeze--out of selected neutral species, including CO and N$_2$, on to the grains and the evolution of the fractional abundances of the major ions, including H$^+$, H$_3^+$ and HCO$^+$. Results are shown for a reference model in which coagulation was neglected and the timescale for collapse is determined by the free--fall time, $\tau _{\rm c} = \tau _{\rm ff}$. The corresponding density profile, $n_{\rm H}(t)$, is also shown in Fig.~\ref{f-o}. As will be seen below, collapse gets under way after a few times $10^5$ yr, and the neutral species then freeze out rapidly on to the grains. Initially, HCO$^+$ is the most abundant molecular ion, but it is overtaken by H$_3^+$ and H$^+$ as freeze--out proceeds. Ultimately, the degree of ionization of the gas is determined by H$_3^+$ under the conditions of this model. 

The adsorption energy of N$_2$ which was used to obtain the results in Fig.~\ref{f-o} was 790~K, which is to be compared with $E_{{\rm CO}}^{\rm ads} = 855$~K (Oberg et al. 2004). A much larger difference between the adsorption energies of N$_2$ and CO is required for there to be a significant difference between the density--dependence of the fractional abundances of N$_2$ and CO, and hence of N$_2$H$^+$ and HCO$^+$. Another possible explanation of this differential effect will be considered below, in the context of a discussion of the nitrogen chemistry.

The effect of allowing the ambipolar diffusion timescale to regulate the collapse is shown in Fig.~\ref{f-o.ad}a. As we shall see below, the timescale for ambipolar diffusion is about an order of magnitude larger than the free--fall time, and so freeze--out occurs at lower values of the gas density, before the collapse is under way. Indeed, in this case, freeze--out occurs at densities which are lower than the values of $n_{\rm H} \gtrsim 3\times 10^4$ cm$^{-3}$ which have been deduced from observations of CO (cf. Bacmann et al.~2002, Tafalla et al.~2004). Thus, the observational data pertaining to CO depletion favour a collapse timescale of order of the free--fall time. The `knee' which is apparent in the curves arises at the fractional abundance at which cosmic ray desorption begins to be significant; this occurs marginally sooner for N$_2$ than for CO, owing to the smaller (by 65~K) adsorption energy of the former species. 

Including coagulation, with the expression of Chokshi et al. (1993) for $v_{\rm crit}(a_{g})$ [equation~(\ref{equ11})], has very little effect on the results, as may be seen by comparing Figs.~\ref{f-o.ad}a and \ref{f-o.ad}b: the `free-fall' results in Fig.~\ref{f-o.ad}a (the upper pair of curves) are almost coincident with the full curves in Figs.~\ref{f-o.ad}b. Even when the (4 times larger) values of $v_{\rm crit}(a_{g})$ are adopted, following Poppe \& Blum (1997), which possibly yields an upper limit to the rate of coagulation, its effect on the evolution of the fractional abundances of the neutral species is not large, as may be seen from Fig.~\ref{f-o.ad}b.

\begin{figure}
\centering
\includegraphics[height=15cm]{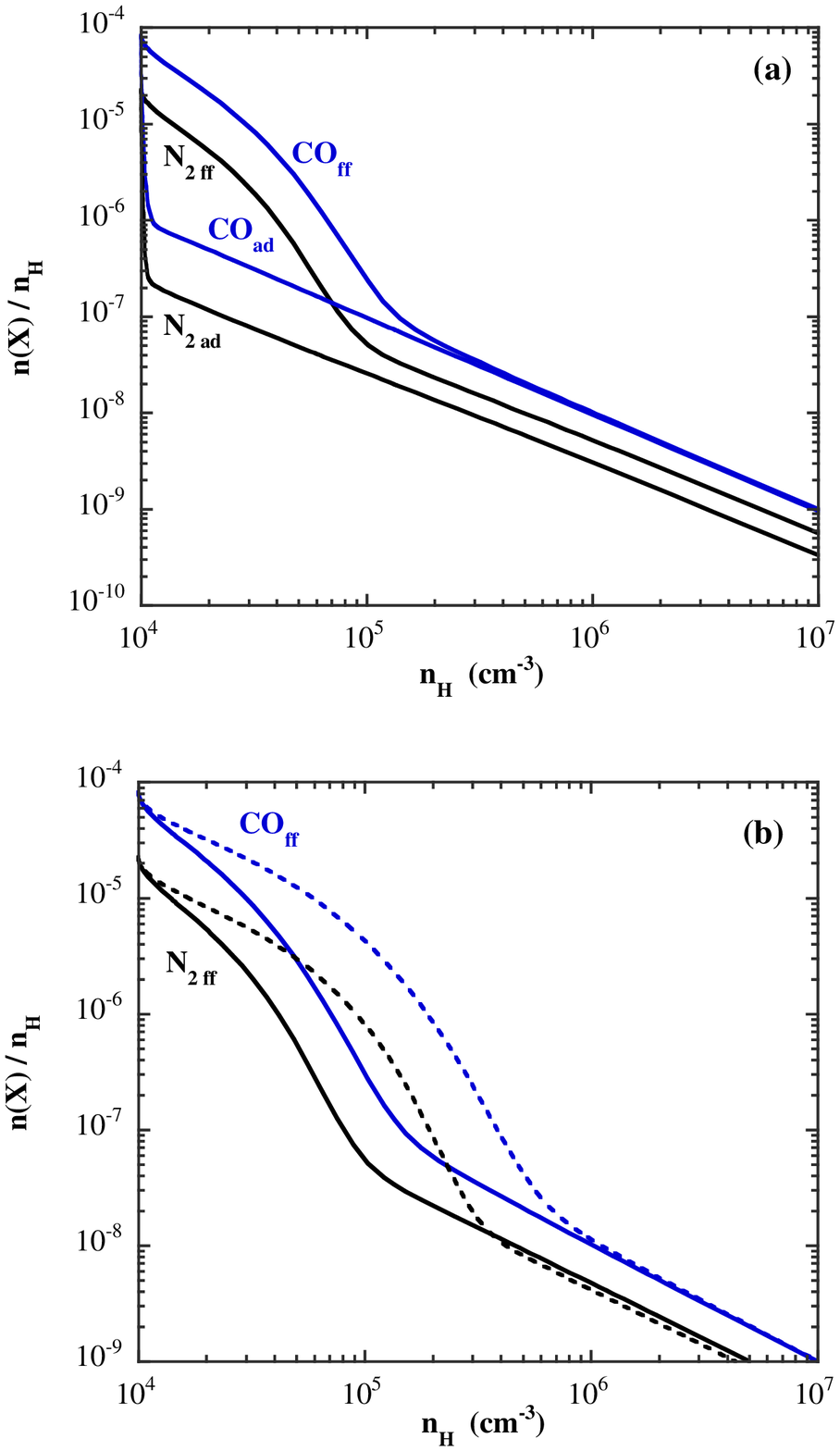}
\caption{The fractional abundances of the exemplary neutrals, N$_2$ and CO, as functions of the gas density, $n_{\rm H}$. Results are shown (a) for the limiting case of no coagulation, with the timescale for collapse determined by either the free--fall time, $\tau _{\rm ff}$, or the ambipolar diffusion time, $\tau _{\rm ad}$; (b) including coagulation, with $v_{\rm crit}(a_{\rm g}) = 0.1\ a_{\rm g}^{-5/6}$ (Chokshi et al. 1993; full curves) or $v_{\rm crit}(a_{\rm g}) = 0.4\ a_{\rm g}^{-5/6}$ (Poppe \& Blum 1997; broken curves). $\Delta v = c_{\rm s}/4$, where $\Delta v$ is the turbulent width and $c_{\rm s}$ is the adiabatic sound speed. The timescale for collapse is determined by the free--fall time.}
\label{f-o.ad}
\end{figure}

\subsection{Timescales}

\begin{figure}
\centering
\includegraphics[height=10cm]{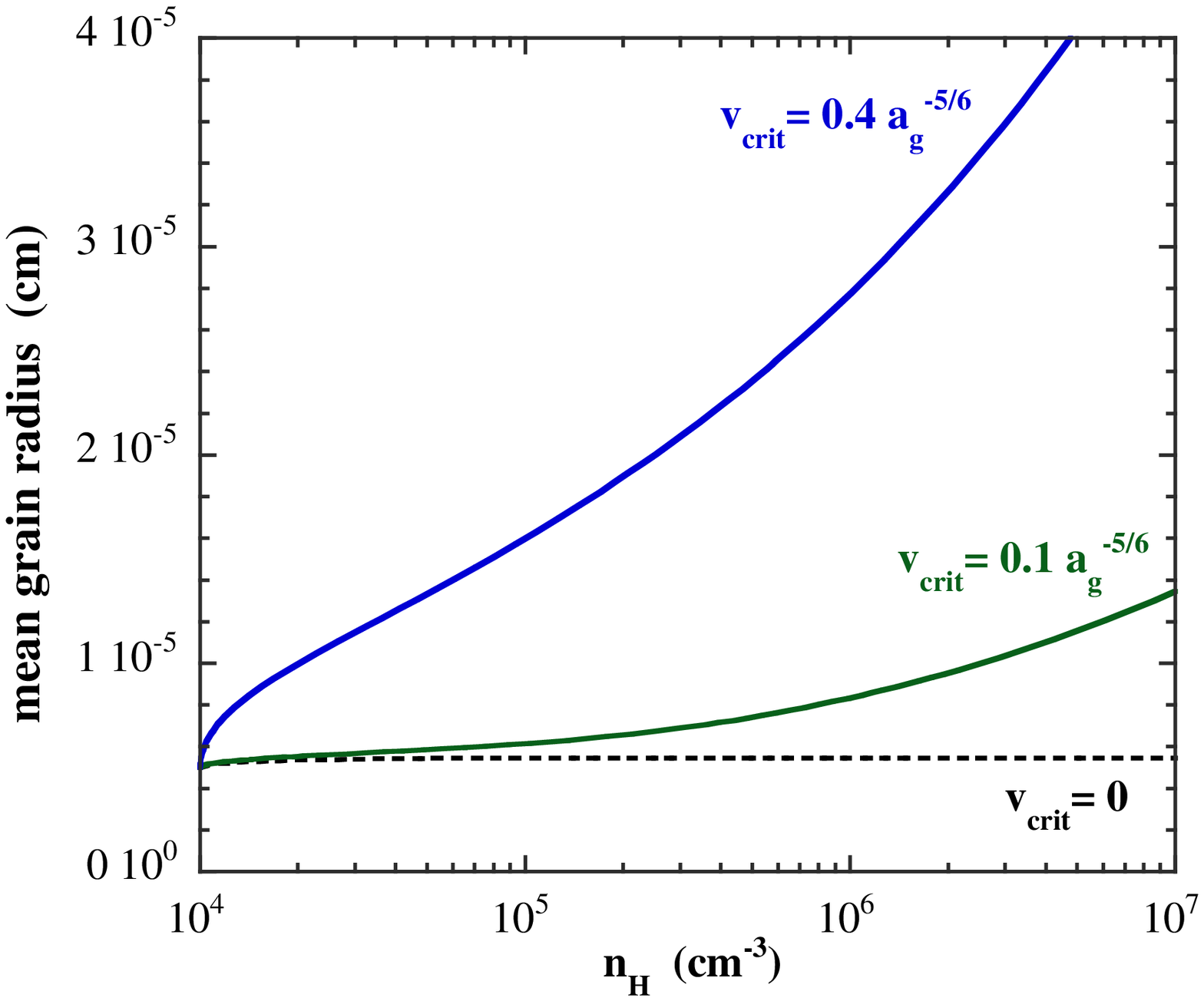}
\caption{The variation of the grain radius, $a_{\rm g}$, as a function of the gas density. Results are shown for the limiting case of no coagulation ($v_{\rm crit} = 0$) and for $v_{\rm crit}(a_{\rm g}) = 0.1\ a_{\rm g}^{-5/6}$ (Chokshi et al. 1993) and $v_{\rm crit}(a_{\rm g}) = 0.4\ a_{\rm g}^{-5/6}$ (Poppe \& Blum 1997); $\Delta v = c_{\rm s}/4$, where $\Delta v$ is the turbulent width and $c_{\rm s}$ is the adiabatic sound speed. The timescale for collapse is determined by the free--fall time, $\tau _{\rm ff}$.}
\label{timescales}
\end{figure}

In Fig.~\ref{timescales}, we plot the grain radius, $a_{\rm g}$, as a function of the gas density, $n_{\rm H}$. Results are shown for the limiting case of no coagulation ($v_{\rm crit} = 0$) and for $v_{\rm crit}(a_{\rm g}) = 0.1\ a_{\rm g}^{-5/6}$ [equation~(\ref{equ11}); Chokshi et al. (1993)] and $v_{\rm crit}(a_{\rm g}) = 0.4\ a_{\rm g}^{-5/6}$ (Poppe \& Blum 1997); $\Delta v = c_{\rm s}/4$, where $c_{\rm s} = 0.243$ km s$^{-1}$ is the adiabatic sound speed. When equation~(\ref{equ11}) is adopted, the initial value of the critical speed for coagulation is $v_{\rm crit} = 0.0262$ km s$^{-1}$, almost an order of magnitude smaller than $c_{\rm s}$. 

Fig.~\ref{timescales} shows that, when no coagulation occurs, the grain radius, $a_{\rm g}$, is practically constant, as the contribution of continuing freeze--out to the size of the grain mantles is small; we have assumed that the grains have a mantle of ices before collapse begins. The rate of coagulation, and hence the rate of increase of the grain radius, $a_{\rm g}$, is seen to be sensitive to the critical speed, $v_{\rm crit}(a_{\rm g})$: increasing the constant of proportionality in equation~(\ref{equ11}) from 0.1 to 0.4, which is in line with the measurements of Poppe \& Blum (1997), has the effect of enhancing significantly the degree of coagulation over the free--fall time. Reducing the turbulent width, $\Delta v$, which  is the FWHM of the Gaussian distribution [equation~(\ref{equA3})], has a similar effect to increasing $v_{\rm crit}$. In our model, the grain opacity per H nucleus, $(n_{\rm g}/n_{\rm H})\pi a_{\rm g}^2$, varies inversely with the grain radius, $a_{\rm g}$, and so the effect of coagulation is to reduce the grain opacity and hence increase the timescale for depletion. The assumption which we have made, that the  grains remain spherical, minimizes their surface area (and hence their geometrical cross section) for a given density of the grain material, which we take to be constant during coagulation. Although these assumptions are not realistic (cf. Ossenkopf and Henning 1994), they allow a lower limit to be placed on the influence of coagulation on freeze--out.

The gravitational free-fall time, $\tau _{\rm ff} = 0.43\times 10^6$~yr for our initial density $n_{\rm H} = 10^4$ cm$^{-3}$, is much smaller than the initial ambipolar diffusion timescales for both positive ions and charged grains, $\tau _{\rm ad}^{\rm ion} = 3.4\times 10^6$~yr and $\tau _{\rm ad}^{\rm grain} = 3.6\times 10^6$~yr. The timescale for freeze--out of the neutral species on to the grains -- the inverse of the adsorption rate -- is given by

\begin{equation}
\tau _{\rm fo} = \left [ n_{\rm g} \pi a_{\rm g}^2 v_{\rm th} S \right ]^{-1}
\label{equ12}
\end{equation}
where 

\begin{displaymath}
v_{\rm th} = \left (\frac {8k_{\rm B}T_{\rm n}}{\pi m_{\rm n}} \right )^{\frac {1}{2}}
\end{displaymath}
is the thermal speed, $T_{\rm n}$ is the kinetic temperature and $m_{\rm n}$ is the mass of the species; $S$ is the sticking coefficient. Taking $T_{\rm n} = 10$~K and $m_{\rm n} = 28\ m_{\rm H}$ (corresponding to CO, as an example) and $S = 1$, we obtain $\tau _{\rm fo} = 2.2\times 10^5$~yr initially, when $a_{\rm g} = 0.05$ $\mu $m and $n_{\rm H} = 10^4$ cm$^{-3}$. Thus, the timescale characterizing freeze--out is comparable with the free--fall timescale and much smaller than that for ambipolar diffusion. The timescale for coagulation is 

\begin{equation}
\tau _{\rm cg} = \left [ n_{\rm g} 4\pi a_{\rm g}^2 v_{\rm rel} \right ]^{-1}
\label{equ13}
\end{equation}
where $v_{\rm rel}$ is the relative collision speed. The maximum possible value of $v_{\rm rel}$ is $v_{\rm crit}$ [cf. equation~(\ref{equ11})]; adopting $v_{\rm rel} = v_{\rm crit}$, we see that

\begin{equation}
\frac {\tau _{\rm fo}} {\tau _{\rm cg}} \propto \frac {v_{\rm crit}} {v_{\rm th}S} 
\label{equ14}
\end{equation}
With the values of the parameters specified above, we obtain $\tau _{\rm cg} = 1.8\times 10^5$~yr initially, which is approximately equal to $\tau _{\rm fo}$. However, $\tau _{\rm cg}$ tends to increase with time, owing to the increase in $a_{\rm g}$ and consequent decrease in $v_{\rm crit}$. As we have assumed the maximum possible value of $v_{\rm rel}$ when evaluating $\tau _{\rm cg}$, we conclude that freeze--out tends to occur more rapidly than coagulation, providing the sticking coefficient $S \approx 1$.

\subsection{Fractional ionization}

The significance of ambipolar diffusion and, in particular, the role of elastic scattering between neutrals and charged grains in protostellar collapse have been recognized for some considerable time (Nakano \& Umebayashi 1980; Shu et al. 1987; Ciolek \& Basu 2000). Whether the timescale for collapse, $\tau _{\rm c}$, is determined by ambipolar diffusion or by gravitational free--fall depends on the fractional ionization of the gas [cf. equation~(\ref{equ9})] and the degree of charge of the grains. 

\begin{figure}
\centering
\includegraphics[width=10cm,height=12cm]{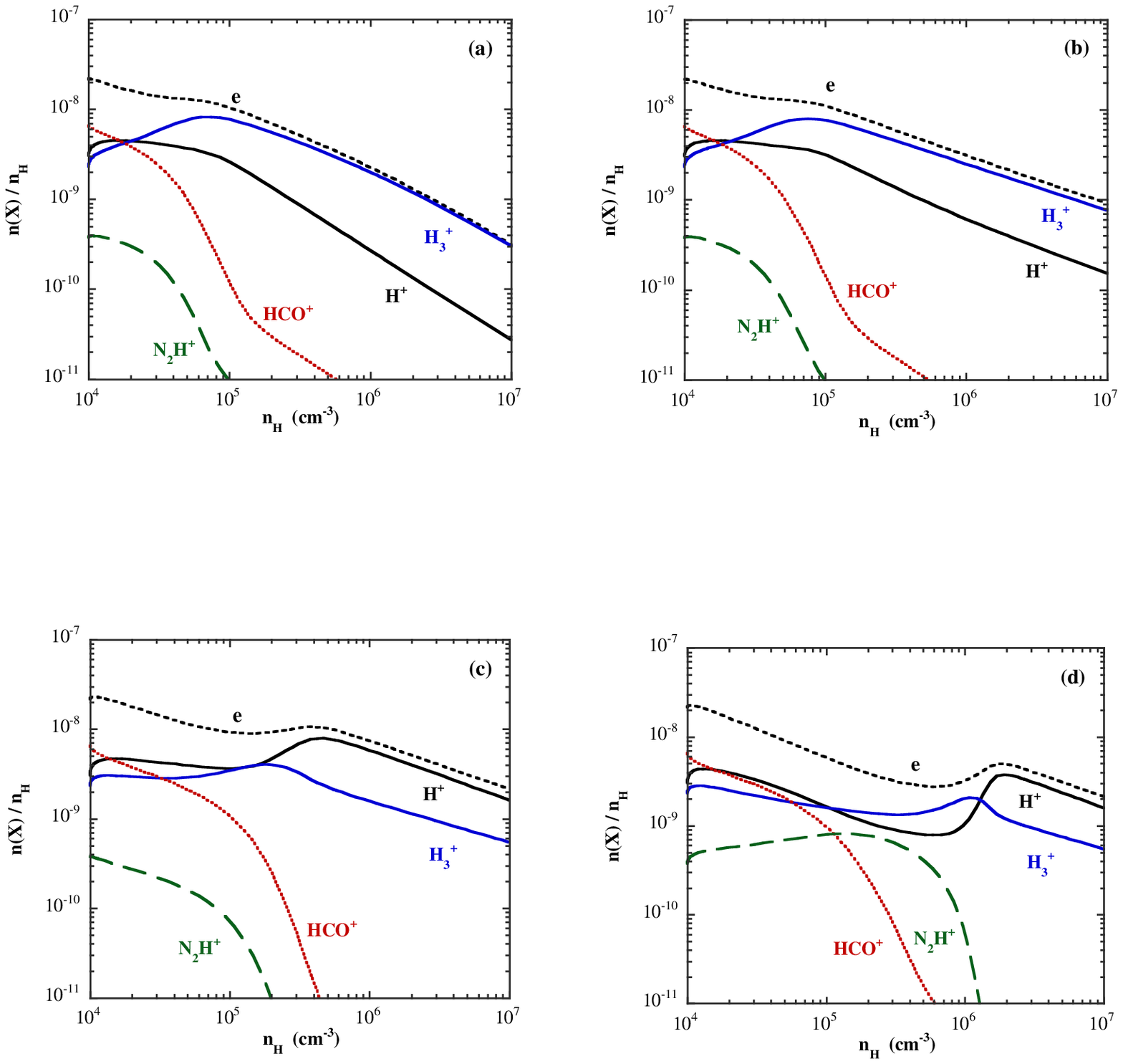}
\caption{The fractional abundances of the specified gas--phase ions and the electrons, as functions of the gas density. Results are shown for (a) the limiting case of no coagulation ($v_{\rm crit} = 0$), (b) $v_{\rm crit}(a_{\rm g}) = 0.1\ a_{\rm g}^{-5/6}$ [equation~(\ref{equ11}); Chokshi et al. (1993)], and (c) $v_{\rm crit}(a_{\rm g}) = 0.4\ a_{\rm g}^{-5/6}$ (Poppe \& Blum 1997); $\Delta v = c_{\rm s}/4$, where $\Delta v$ is the turbulent width and $c_{\rm s}$ is the adiabatic sound speed. The results in panel (d) were obtained on reducing the value of the sticking coefficient for molecular nitrogen to $S({\rm N}_2) = 0.1$ and should be compared with those in panel (c), where $S({\rm N}_2) = 1$. In all cases, the timescale for collapse is determined by the free--fall time, $\tau _{\rm ff}$.}
\label{vst}
\end{figure}

At early times, S$^+$ is the most abundant atomic ion, and its abundance is maintained by charge transfer of H$^+$ to S; the source of H$^+$ is cosmic ray ionization of hydrogen. When freeze--out of the neutrals on to the grains occurs, the fractional abundance of H$^+$ tends to rise as losses through charge transfer reactions (with S and other neutral species with ionization potentials less than that of H) decrease.  This phenomenon is illustrated in Fig.~\ref{vst}, where the fractional abundances of H$^+$ and a selection of other ions are plotted as functions of the gas density, $n_{\rm H}$. The significance of coagulation may be seen by comparing Fig.~\ref{vst}a, in which coagulation is neglected, with Figs.~\ref{vst}b and c, in which coagulation is included.

The recombination of H$^+$ with free electrons is a radiative process and hence slow. Consequently, H$^+$ ions recombine with electrons predominantly on the surfaces of negatively charged grains. On the other hand, the recombination of H$_3^+$ with free electrons is a non--radiative (dissociative) process and intrinsically fast. Using the parameters adopted in our model, the ratio $R$ of the rate (s$^{-1}$) of dissociative recombination of H$_3^+$ in the gas phase and its rate of recombination on the surfaces of the (predominantly) negatively charged grains is 

\begin{displaymath}
R \approx 2\times 10^{11}x_{\rm e} a_{\rm g}^2 
\end{displaymath}
where $x_{\rm e} = n_{\rm e}/n_{\rm H}$ and $a_{\rm g}$ is expressed in $\mu $m. Thus, when $a_{\rm g} = 0.05$ $\mu $m, $R = 1$ is attained when $x_{\rm e} \approx 2\times 10^{-9}$; then, H$_3^+$ recombines at equal rates on the surfaces of negatively charged dust grains and in the gas phase. Referring to Fig.~\ref{vst}a, we see that this point is reached when $n_{\rm H} \approx 10^6$ cm$^{-3}$. At lower densities, dissociative recombination predominates, whereas, at higher densities, recombination on the surfaces of negative grains predominates, and the fractional abundances of H$_3^+$ and H$^+$ vary similarly with $n_{\rm H}$. The ratio of the fractional abundances of H$_3^+$ and H$^+$ 

\begin{displaymath}
n({\rm H}_3^+)/n({\rm H}^+) \approx 18/(1 + R)
\end{displaymath}
for the parameters adopted in our model; this expression agrees well with the numerical results presented in Fig.~\ref{vst}a in the high density limit, where $R \rightarrow 0$. [We recall that, owing to the absence of deuterated isotopes in the present model, the number density of ``H$_3^+$'' should be understood as $n({\rm H}_3^+) + n({\rm H}_2{\rm D}^+) + n({\rm HD}_2^+) + n({\rm D}_3^+)$.]

\begin{figure}
\centering
\includegraphics[height=12cm]{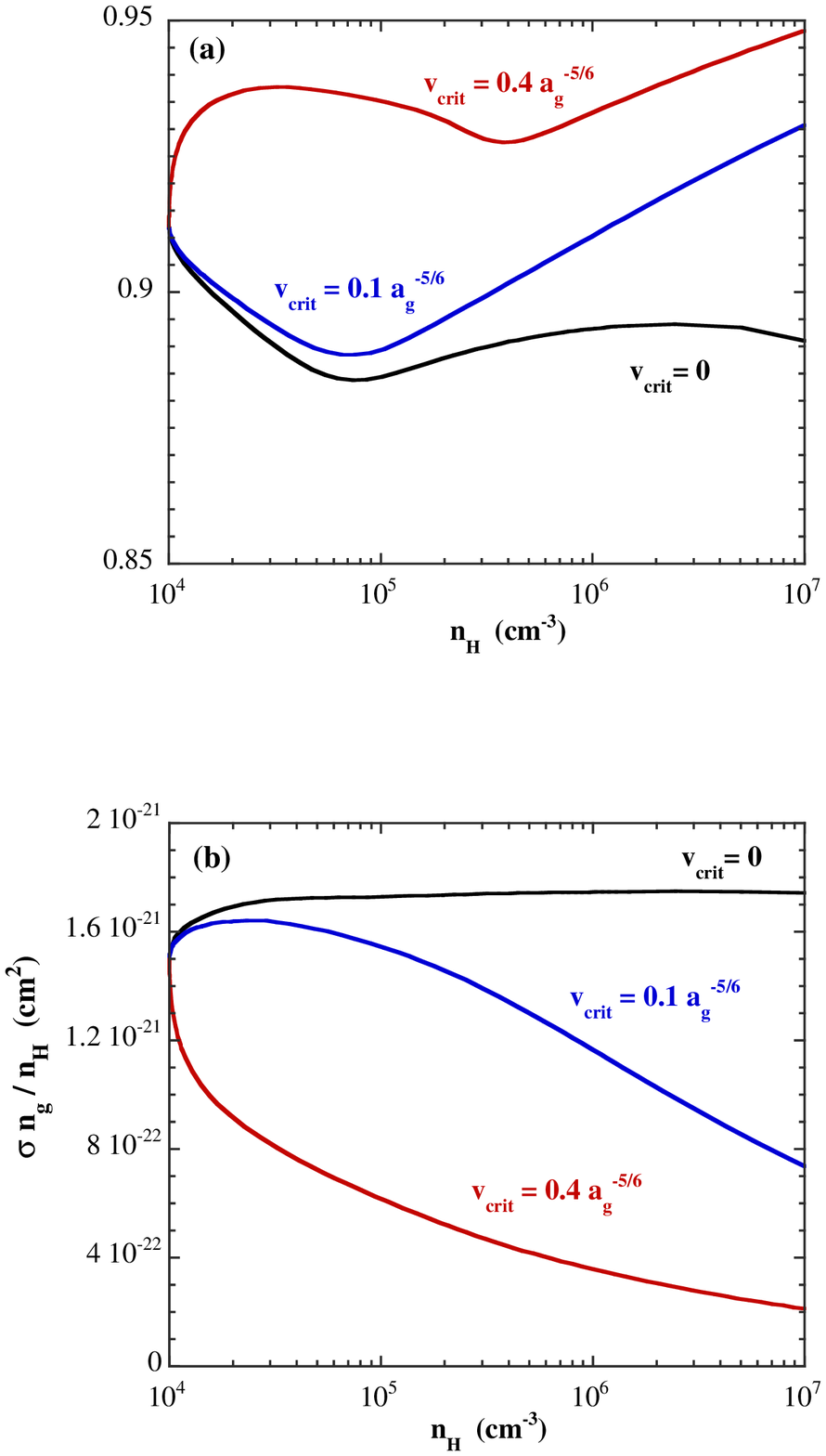}
\caption{(a) The fraction and (b) the opacity per H nucleus, $(n_{\rm g}^-/n_{\rm H})\pi a_{\rm g}^2$, of negatively charged grains, as functions of the gas density, $n_{\rm H}$. Results are shown for the limiting case of no coagulation ($v_{\rm crit} = 0$), for $v_{\rm crit}(a_{\rm g}) = 0.1\ a_{\rm g}^{-5/6}$ [equation~(\ref{equ11}); Chokshi et al. (1993)], and for $v_{\rm crit}(a_{\rm g}) = 0.4\ a_{\rm g}^{-5/6}$ (Poppe \& Blum 1997); $\Delta v = c_{\rm s}/4$, where $\Delta v$ is the turbulent width and $c_{\rm s}$ is the adiabatic sound speed. The timescale for collapse is determined by the free--fall time, $\tau _{\rm ff}$.}
\label{ions}
\end{figure}

In the case where there is no coagulation, as in Fig.~\ref{vst}a, the grain opacity per H nucleus, $(n_{\rm g}/n_{\rm H})\pi a_{\rm g}^2$, remains practically constant as the collapse proceeds; cf. Fig.~\ref{ions}b. The H$^+$ ions are removed (by recombination with electrons on the surfaces of negatively charged grains) on a timescale which is inversely proportional to $n_{\rm g}\pi a_{\rm g}^2$ and hence to $n_{\rm H}$. For $n_{\rm H} \lesssim 10^6$ cm$^{-3}$, H$_3^+$ ions are removed predominantly by dissociative recombination with free electrons, on a timescale which is inversely proportional to approximately $n_{\rm H}^{0.5}$. As a consequence, the fractional abundance of H$_3^+$ ions initially increases relative to H$^+$ as $n_{\rm H}$ increases, and the case of `no coagulation' is characterized by $n({\rm H}_3^+) > n({\rm H}^+)$ at high gas densities. On the other hand, when coagulation occurs (assuming $\Delta v = c_{\rm s}/4$ in Fig.~\ref{vst}), $(n_{\rm g}/n_{\rm H})\pi a_{\rm g}^2$ decreases as $n_{\rm H}$ increases (cf. Fig.~\ref{ions}b), and the rate of recombination of the H$^+$ ions on grains is reduced. Owing to the higher fractional abundance of the free electrons, $n({\rm H}_3^+) < n({\rm H}^+)$ at high gas densities when $v_{\rm crit}(a_{\rm g}) = 0.4\ a_{\rm g}^{-5/6}$. These results are summarized in Figs.~\ref{vst}a--c.

The results in Fig.~\ref{vst} show that the ratio $n({\rm H}_3^+)/n({\rm H}^+)$ falls substantially when coagulation is introduced: grain growth favours H$^+$. On the other hand, the fact that H$_2$D$^+$ and D$_2$H$^+$ have been observed in pre--protostellar objects (Caselli et al.~2003, Vastel et al.~2004) suggests that the deuterated forms of H$_3^+$ may be the dominant gas--phase ions. If so, the present calculations imply that coagulation is not significant in reducing the grain opacity per H nucleus during the early phases of collapse. 

The final panel (d) in Fig.~\ref{vst} shows the effect of reducing the value of the sticking coefficient for molecular nitrogen to $S({\rm N}_2) = 0.1$ and should be compared with those in panel (c), where $S({\rm N}_2) = 1$, the default value for all species which freeze out; the results in Fig.~\ref{vst}d are considered below.

A novelty of the present study is the calculation of the grain charge distribution, in parallel with the chemistry and the dynamics; the charge distribution is significant because H$^+$ ions recombine predominantly on the surfaces of negatively charged grains, at a rate which is determined by their opacity per H nucleus, $(n_{\rm g}^-/n_{\rm H})\pi a_{\rm g}^2$. The higher fractional abundance of free electrons at high gas densities, owing to coagulation (cf. Fig.~\ref{vst}), leads to an increase in the fraction of the grains which are negatively charged; see Fig.~\ref{ions}. The fraction of positively charged grains always remains small. When there is no coagulation, the fraction of neutral grains remains above 10\% as the collapse proceeds, whereas, for the maximum rate of coagulation considered in Fig.~\ref{ions}, approximately 95\% of the grains are negatively charged by the time that $n_{\rm H} = 10^7$ cm$^{-3}$. Nonetheless, the opacity per H nucleus of the negatively charged grains, $(n_{{\rm g}^-}/n_{\rm H})\pi a_{\rm g}^2$, decreases as $v_{\rm crit}$ increases, owing to the reduction in the {\it total} grain opacity per H nucleus as coagulation proceeds: see Fig.~\ref{ions}b.

\subsection{Nitrogen chemistry}

One problem which all models of the chemistry of prestellar
cores have to solve is the variation of the fractional 
abundances of nitrogen--containing compounds with density.
Observations indicate that, whilst most molecules deplete
on to grain surfaces at densities of a few times $10^4$ cm$^{-3}$,
species containing nitrogen, such as NH$_3$ and N$_2$H$^+$, survive
to densities which are an order of magnitude larger (e.g. Tafalla et al.
2004, Belloche \& Andr\'{e} 2004, Bergin et al. 2002).  Chemical modellers have attempted to explain this behaviour by assuming lower values of the adsorption energy for N$_2$ than for CO and other abundant neutral molecules; 
this results in cosmic ray--induced or direct thermal desorption of N$_2$ 
being more efficient than for CO. Aikawa et al. (1997, 2001) and Bergin et al. (2002) give $E_{{\rm N}_2}^{\rm ads} = 750$~K as the adsorption energy of N$_2$ and $E_{{\rm CO}}^{\rm ads} = 960$~K as the adsorption energy of CO, on CO ice. On the other hand, recent measurements (Oberg et al. 2004) indicate that the adsorption energies of N$_2$ to N$_2$ and CO to CO ice are $E_{{\rm N}_2}^{\rm ads} = 790$~K and $E_{{\rm CO}}^{\rm ads} = 855$~K; we have adopted these values in our calculations
but considered the implications of using larger values of $(E_{{\rm CO}}^{\rm ads} - E_{{\rm N}_2}^{\rm ads})$. We have concluded that the approximately 10\% difference between the binding energies of CO and N$_2$, as measured by Oberg et al. (2004), 
is insufficient to account for the observed differences in fractional abundances, to which reference was made above. We note that, in the present context, it is the {\it ratio} between the rates of desorption of CO and N$_2$ which is significant and determined by the difference in their adsorption energies; the {\it absolute} rates are less crucial.

An possible explanation of the observations is that there is a difference
between the sticking coefficients of CO and one of the principal forms of gas--phase nitrogen (N or N$_2$). Accordingly, we have carried out calculations in which the sticking coefficient, $S$, of either N or N$_2$ was set
equal to 0.1, whilst retaining $S = 1$ for all other species; the results of the calculations in which $S({\rm N}_2) = 0.1$ are shown in Fig.~\ref{vst}. In Fig.~\ref{vst}c and Fig.~\ref{vst}d, 
we compare the free--fall models, including coagulation and 
with $S({\rm N}_2) = 1$ or $S({\rm N}_2) = 0.1$, respectively.  In the case of $S({\rm N}_2) = 0.1$, the fractional abundance of N$_2$H$^+$, which traces N$_2$, decreases rapidly at densities
$n_{\rm H} \gtrsim 10^6$ cm$^{-3}$, which is roughly an order of magnitude higher than for the corresponding fall--off of HCO$^+$. The fractional abundance of N$_2$H$^+$ attains values of several times $10^{-10}$, which are, in fact, larger than the observed values. However, the effect upon
the evolution of the major ions, H$^+$ and H$_3^+$, is not large,
and N$_2$H$^+$ never becomes a major ion. We conclude that
a small sticking coefficient for N$_2$ could explain 
the observations; but this assumption is clearly ad hoc and needs experimental verification. 

The other possibility is that the sticking coefficient of atomic nitrogen is small, $S({\rm N}) = 0.1$. This assumption has the consequence that a substantial fraction of the nitrogen remains in the gas phase, but it does not result in a significant enhancement of the abundance of N$_2$H$^+$, relative to the case in which all the sticking coefficients $S = 1$. The reason for this result is to be found in the gas--phase chemistry of nitrogen. The paths to N$_2$H$^+$ and NH$_3$ start with the neutral--neutral reactions N(OH, H)NO and NO(N, O)N$_2$ (Pineau des For\^{e}ts et al. 1990), whose rates depend on the fractions of N and OH in the gas phase. However, these reactions are so slow that an increase in the initial gas--phase abundance of N has little effect on the abundances of N$_2$H$^+$ or NH$_3$ {\it on the timescale of the collapse}. Only when the abundance of N$_2$ is increased directly, by reducing its sticking coefficient or its adsorption energy, are the abundances of N$_2$H$^+$ and NH$_3$ enhanced significantly.

\subsection{Validation of previous results}

\begin{figure}
\centering
\includegraphics[height=15cm]{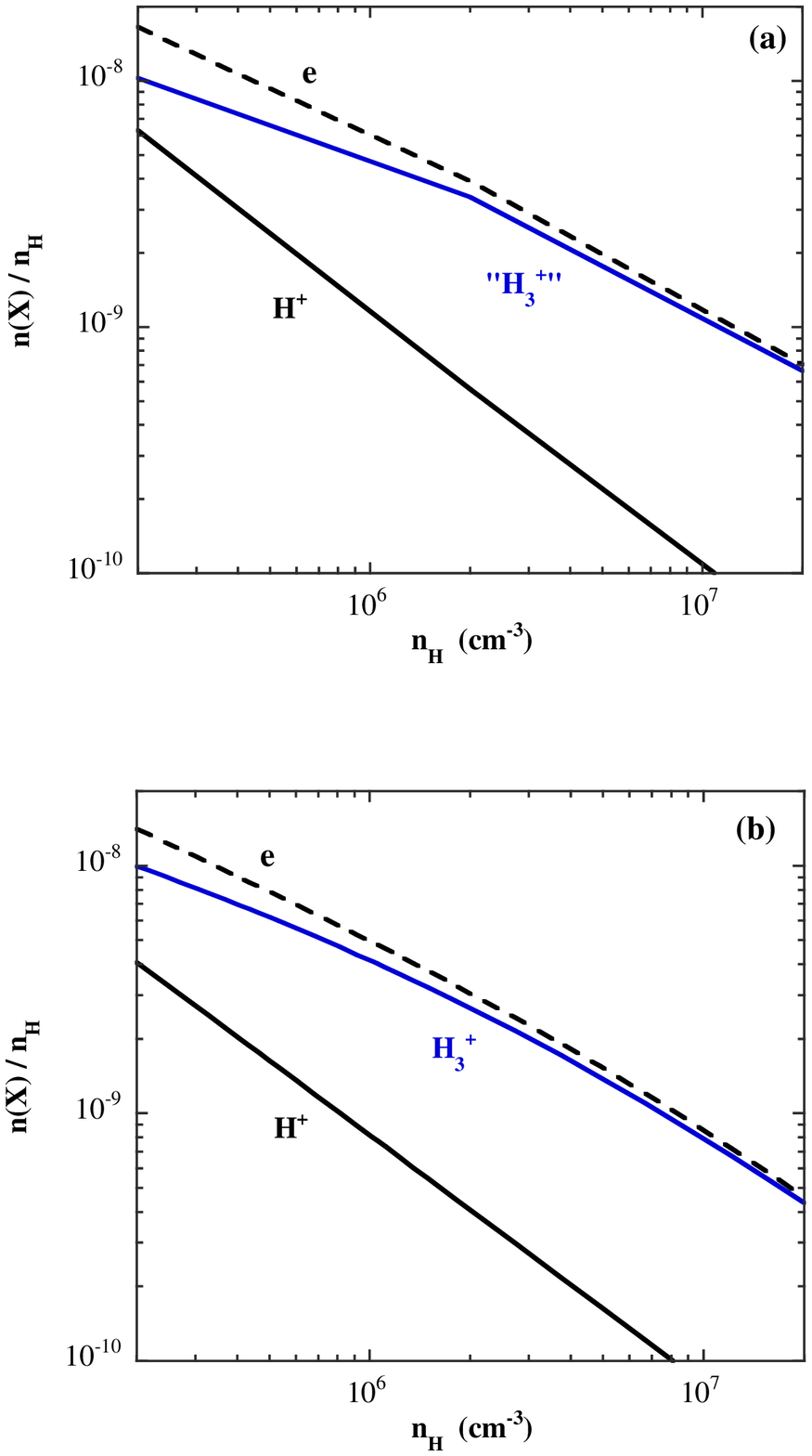}
\caption{A comparison of the fractional abundances of the principal ions, (a) as computed by Walmsley et al. (2004), assuming steady--state, and (b) as predicted by the present dynamical model; $a_{\rm g} = 0.05$ $\mu $m and, in this case only, $\zeta = 3\times 10^{-17}$ s$^{-1}$, in order to be able to compare directly with Walmsley et al. (2004). [``H$_3^+$'' denotes $n({\rm H}_3^+) + n({\rm H}_2{\rm D}^+) + n({\rm HD}_2^+) + n({\rm D}_3^+)$.]}
\label{validation}
\end{figure}

In our previous calculations of complete depletion in pre--protostellar cores (Walmsley et al. 2004; Flower et al. 2004), we computed abundances in steady--state, neglecting dynamical effects but including deuteration reactions. Although we do not currently consider deuteration reactions in the phase of `free--fall', it is possible to compare the fractional ionization of the gas and the abundances of the major ions, H$^+$ and H$_3^+$, over the range of densities considered previously ($2\times 10^5 \le n_{\rm H} \le 2\times 10^7$ cm$^{-3}$). In order to make the appropriate comparison with our previous work, we plot, in Fig.~\ref{validation},  $n({\rm H}_3^+)/n_{\rm H}$ from the present calculations, neglecting coagulation, and $[n({\rm H}_3^+) + n({\rm H}_2{\rm D}^+) + n({\rm HD}_2^+) + n({\rm D}_3^+)]/n_{\rm H}$ from our previous study (Walmsley et al. 2004).

As may be seen from Fig.~\ref{validation}, the agreement between the two calculations is good. Although the abundances of atomic ions such as Fe$^+$ and, especially, S$^+$ are initially significant (i.e when $n_{\rm H} \approx 10^4$ cm$^{-3}$), recombination with electrons, on the surfaces of grains, neutralizes the atomic ions. By the time that a density of $n_{\rm H} \approx 10^5$ cm$^{-3}$ has been reached, the principal ions are H$_3^+$ and H$^+$, as predicted by our earlier calculations, in steady--state. Thus, even when the collapse is as fast as in the case of `free--fall', the assumption of a steady--state is a good approximation.

\section{Concluding remarks}

We have considered the early stages of isothermal collapse of a pre--protostellar core. The initial conditions were taken to be those appropriate to dense molecular clouds, with significant fractions of the heavy elements being depleted into the cores or the ice mantles of dust grains. The time for collapse is taken equal to the timescale which characterizes either free--fall or ambipolar diffusion; the latter is almost an order of magnitude larger than the former. 

The processes of further freeze--out on to the interstellar grains and of coagulation of the grains were included when calculating the chemical evolution of the medium in the course of its collapse. Species containing heavy elements freeze on to the grains in the early phases of the collapse. When ambipolar diffusion determines the timescale for collapse, freeze--out occurs at densities which are lower than the values which have been deduced from observations of CO (cf. Bacmann et al.~2002, Tafalla et al.~2004); we conclude that the observational data favour a collapse timescale of order of the free--fall time, rather than the (larger) ambipolar diffusion time. In this case, coagulation is unlikely to be sufficiently rapid to influence the rate of freeze--out.

The rate of grain coagulation is determined by the turbulent velocity, $\Delta v$, of the medium. In the limit of $\Delta v << c_{\rm s}$, the rate of coagulation is negligible. On the other hand, when proceeding at close to its maximum rate, coagulation has important consequences for the degree of ionization and the ionic composition of the medium. At high gas densities, the major ions are H$^+$ and H$_3^+$, and their relative abundance depends on the rate of coagulation, through its effect on the grain opacity per H nucleus, $(n_{\rm g}/n_{\rm H})\pi a_{\rm g}^2$; this is because H$^+$ ions recombine on the surfaces of negatively charged grains, whereas H$_3^+$ ions also recombine dissociatively with free electrons in the gas phase. If coagulation is slow or absent, $n({\rm H}_3^+) > n({\rm H}^+)$, whereas, if coagulation is rapid, $n({\rm H}_3^+) < n({\rm H}^+)$ at high densities. The observations of H$_2$D$^+$ and D$_2$H$^+$ in pre--protostellar objects suggest that H$_3^+$ and its deuterated forms are likely to be the dominant ions, in which case coagulation is slow or absent.

We consider that it is important to calculate the grain charge distribution, in parallel with the chemistry and the dynamics, because H$^+$ ions recombine predominantly on the surfaces of negatively charged grains. Grain coagulation results in an increase in the fraction of electrons which remain in the gas phase at high densities and to an increase in the fraction of the grains which are negatively charged; the fraction of positively charged grains is always small.

Observations of pre--protostellar cores indicate that nitrogen--containing species, such as NH$_3$ and N$_2$H$^+$, survive to densities which are an order of magnitude larger than those at which most molecules freeze on to grain surfaces (Tafalla et al. 2004, Belloche \& Andr\'{e} 2004, Bergin et al. 2002). This differential depletion was believed to be due to the adsorption energy of N$_2$ being significantly lower than those of other molecules, such as CO (see Bergin \& Langer 1997); but recent measurements of the adsorption energies of CO and N$_2$ (Oberg et al. 2004) do not support this interpretation. In order to solve this problem, a mechanism must be found which results in nitrogen--containing species remaining in the gas phase at densities above 10$^5$ cm$^{-3}$; the gas--phase chemistry must also be adequately described under these conditions. We have considered the consequences of reducing the sticking coefficient of N or N$_2$ to values of the order of 0.1. Only when $S({\rm N}_2) = 0.1$, whilst retaining $S = 1$ for all other species, are the abundances of both NH$_3$ and N$_2$H$^+$ enhanced, relative to CO, during the course of the collapse. However, we recognize that the assumption of $S({\rm N}_2) \approx 0.1$ is ad hoc and needs experimental verification.

\appendix

\section{The rate coefficient for grain coagulation}

We assume that the velocity distribution of the population of grains is Gaussian and  consider a collision between two grains, 1 and 2. The $x$-component of the velocity distribution of grain 1 is given by

\begin{equation}
\phi (v_{x_1}) = \frac {h_1}{\pi ^{\frac {1}{2}}} {\rm e}^{-h_1^2 v_{x_1}^2}
\label{equA1}
\end{equation}
such that

\begin{equation}
\int _{-\infty }^{\infty }\phi (v_{x_1}) {\rm d}v_{x_1} = 1
\label{equA2}
\end{equation}
Thus the probability that grain 1 has cartesian velocity components in the ranges $v_{x_1} \rightarrow v_{x_1} + {\rm d}v_{x_1}$, $v_{y_1} \rightarrow v_{y_1} + {\rm d}v_{y_1}$, $v_{z_1} \rightarrow v_{z_1} + {\rm d}v_{z_1}$ is

\begin{displaymath}
p(v_{x_1}, v_{y_1}, v_{z_1})\ {\rm d}v_{x_1}\ {\rm d}v_{y_1}\ {\rm d}v_{z_1} \equiv \phi (v_{x_1})\ \phi (v_{y_1})\ \phi (v_{z_1})\ {\rm d}v_{x_1}\ {\rm d}v_{y_1}\ {\rm d}v_{z_1}
\end{displaymath}
\begin{equation}
= \left (\frac {h_1}{\pi ^{\frac {1}{2}}} \right )^3 {\rm e}^{-h_1^2 v_{1}^2}\ {\rm d}v_{x_1}\ {\rm d}v_{y_1}\ {\rm d}v_{z_1}
\label{equA3}
\end{equation}
where $v_{1}^2 = v_{x_1}^2 + v_{y_1}^2 + v_{z_1}^2$ and we have assumed the velocity distribution to be isotropic. An expression completely analogous to (\ref{equA3}) holds for grain 2. Following Margenau \& Murphy (1956), it may be shown that the probability of grain 1 having an $x$-component of velocity in the range $v_{x_1} \rightarrow v_{x_1} + {\rm d}v_{x_1}$ and grain 2 having an $x$-component of velocity in the range $v_{x_2} \rightarrow v_{x_2} + {\rm d}v_{x_2}$ is

\begin{equation}
\phi (v_{x_1})\ \phi (v_{x_2})\ {\rm d}v_{x_1}\ {\rm d}v_{x_2} = \frac {H}{\pi ^{\frac {1}{2}}} {\rm e}^{-H^2 v_{x}^2}\ {\rm d}v_{x}
\label{equA4}
\end{equation}
where

\begin{equation}
\frac {1}{H^2} = \frac {1}{h_1^2} + \frac {1}{h_2^2} = \frac {2}{h^2}
\label{equA5}
\end{equation}
when $h_1 = h_2 \equiv h$; $v_{x} = v_{x_1} - v_{x_2}$ is the {\it relative} $x$-component of velocity of the two grains. Analogous expressions hold for the $y$- and $z$-components of velocity. Thus, the probability of finding the velocity components of the grains within the ranges specified above is

\begin{displaymath}
p(v_{x_1}, v_{y_1}, v_{z_1})\ p(v_{x_2}, v_{y_2}, v_{z_2})\ {\rm d}v_{x_1}\ {\rm d}v_{y_1}\ {\rm d}v_{z_1}\ {\rm d}v_{x_2}\ {\rm d}v_{y_2}\ {\rm d}v_{z_2} 
\end{displaymath}
\begin{displaymath}
= \left (\frac {H}{\pi ^{\frac {1}{2}}} \right )^3 {\rm e}^{-H^2 v^2}\ {\rm d}v_{x}\ {\rm d}v_{y}\ {\rm d}v_{z}
\end{displaymath}
\begin{equation}
\equiv P(v_{x}, v_{y}, v_{z})\ {\rm d}v_{x}\ {\rm d}v_{y}\ {\rm d}v_{z}
\label{equA6}
\end{equation}
where ${\bf v} = {\bf v_1} - {\bf v_2}$ is the relative velocity of the grains 1 and 2. As the distributions are isotropic,

\begin{equation}
{\rm d}v_{x}\ {\rm d}v_{y}\ {\rm d}v_{z} = 4\pi v^2\ {\rm d}v
\label{equA7}
\end{equation}
Defining

\begin{displaymath}
{\cal P}(v) \equiv 4\pi P(v_{x}, v_{y}, v_{z})
\end{displaymath}
\begin{equation}
= 4\pi \left (\frac {H}{\pi ^{\frac {1}{2}}} \right )^3 {\rm e}^{-H^2 v^2}
\label{equA8}
\end{equation}
we have the overall normalization condition

\begin{equation}
\int _{0}^{\infty }{\cal P} (v) v^2\ {\rm d}v = 1
\label{equA9}
\end{equation}

The rate coefficient for coagulation in grain--grain collisions is 

\begin{equation}
\langle \sigma v \rangle = 4\pi a_{\rm g}^2 \int _0^{v_{\rm crit}} v {\cal P} (v) v^2 {\rm d}v
\label{equA10}
\end{equation}
where $\sigma = 4\pi a_{\rm g}^2$ is the cross section for collisions between identical grains; this is the expression for the geometrical cross section, assuming the grains to be spherical, which minimizes their surface area (and effective cross section) for a given volume. Clearly, non--spherical grains would have a larger cross section.  The neglect of the Coulomb and polarization potentials, which arise when at least part of the population of grains is charged, may be seen to be justified as follows.

\begin{figure}
\centering
\includegraphics[height=10cm]{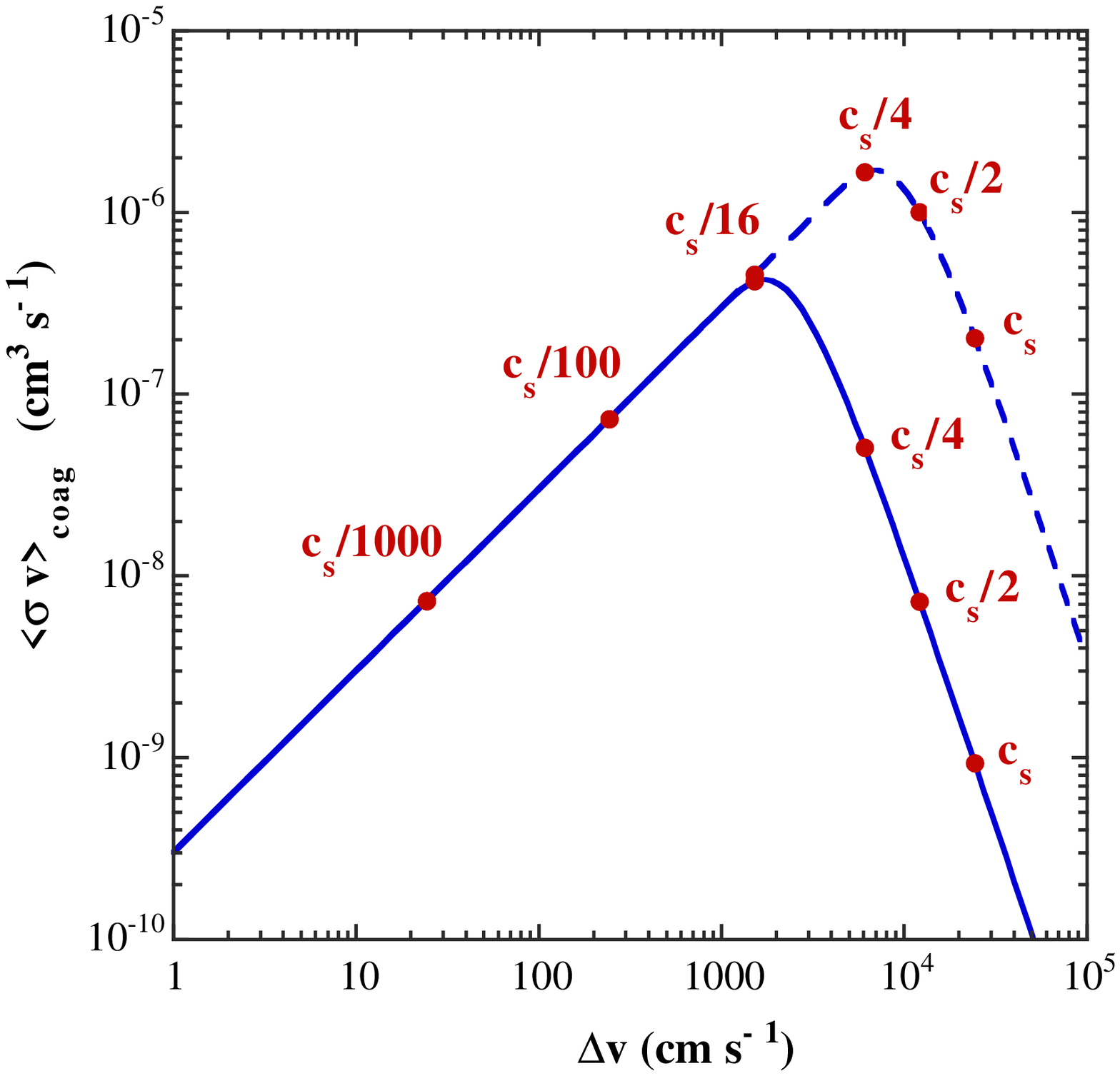}
\caption{The variation of the rate coefficient for coagulation, $\langle \sigma v \rangle $, with the turbulent width, $\Delta v$, for $a_{\rm g} = 0.05$ $\mu $m and  $v_{\rm crit}(a_{\rm g}) = 0.1\ a_{\rm g}^{-5/6}$ (Chokshi et al. 1993; full curve) or $v_{\rm crit}(a_{\rm g}) = 0.4\ a_{\rm g}^{-5/6}$ (Poppe \& Blum 1997; broken curve). Also shown are fractions of the adiabatic sound speed, $c_{\rm s}$, evaluated at $T = 10$~K.}
\label{sigmav}
\end{figure}

The grain population consists of neutral and singly charged grains; of the latter,  negatively charged grains are by far the most abundant. Consider the long--range Coulomb interaction between two grains, each with a single negative charge; this repulsive potential is given by

\begin{displaymath}
V_{\rm Coul} = \frac {e^2}{r},
\end{displaymath}
where $r$ is the separation of the grain centres; $V_{\rm Coul}$ has a maximum value of $e^2/(2a_{\rm g})$, when the grains are in contact. The kinetic energy of a grain, in the centre of mass system, is $m_{\rm g}v_{\rm g}^2/4$, where

\begin{displaymath}
m_{\rm g} = \frac {4}{3}\pi a_{\rm g}^3 \rho _{\rm g}
\end{displaymath}
is the mass of a grain (and $m_{\rm g}/2$ is the reduced mass, for collisions between identical grains); $v_{\rm g}$ is the relative collision speed. Taking $a_{\rm g} = 0.05 \mu$m, which is the minimum (initial) value in the calculations reported in this paper, and $\rho _{\rm g} = 2$ g cm$^{-3}$, we obtain $m_{\rm g} = 1.0\times 10^{-15}$ g. The relative kinetic energy of the grains is greater than or equal to the repulsive Coulomb potential when

\begin{displaymath}
v_{\rm g} \ge \left (\frac {3}{2\pi \rho _{\rm g}} \right )^{\frac {1}{2}} \frac {e}{a_{\rm g}^2} 
\end{displaymath}
i.e. when $v_{\rm g} \gtrsim 10$ cm s$^{-1}$ for the parameters above. On the other hand, the Gaussian velocity distribution of the grains has a FWHM of $\Delta v = c_{\rm s}/4 \approx 6000$ cm s$^{-1}$ in the calculations reported here; see Fig.~\ref{sigmav}. Thus, only a very small section of the velocity distribution, with $v_{\rm g} \lesssim 10$ cm s$^{-1}$ $<< \Delta v$, is affected significantly by the Coulomb interaction. The polarization potential between the grains is of shorter range and attractive; its asymptotic form is

\begin{displaymath}
V_{\rm pol} = -\frac {e^2 a_{\rm g}^3}{2r^4},
\end{displaymath}
where $a_{\rm g}^3$ is the polarizability of a grain of radius $a_{\rm g}$. Taking into account this interaction would tend to enhance the cross section for coagulation of a neutral and a charged grain; but arguments analogous to those immediately above show that the effect would be negligible for the velocity distribution that we have adopted.

Following Chokshi et al. (1993), we assume that the grains stick (coagulate) if the relative collision speed $v \le v_{\rm crit}$. The critical speed, $v_{\rm crit} \propto a_{\rm g}^{-5/6}$, according to Chokshi et al. (1993); the constant of proportionality depends on the grain composition.

Using (\ref{equA8}), the integral in equation~(\ref{equA10}) can be evaluated, yielding

\begin{displaymath}
\langle \sigma v \rangle = \sigma \left (\frac {2}{\pi \ {\rm ln}\ 2}\right )^{\frac {1}{2}} \Delta v \left\{ 1 - \left [1 + 2\ {\rm ln}\ 2\left (\frac {v_{\rm crit}}{\Delta v} \right )^2 \right ] 
{\rm exp} \left [-2\ {\rm ln}\ 2\left (\frac {v_{\rm crit}}{\Delta v} \right )^2 \right ]\right\}
\end{displaymath}
\begin{equation}
= \sigma \ 0.9584\ \Delta v \left\{ 1 - \left [1 + 1.3863\left (\frac {v_{\rm crit}}{\Delta v} \right )^2 \right ] {\rm exp} \left [-1.3863\left (\frac {v_{\rm crit}}{\Delta v} \right )^2 \right ]\right\}
\label{equA11}
\end{equation}
where we have substituted $H^2 = h^2/2$ and defined

\begin{equation}
h^2 = \frac {4\ {\rm ln}\ 2}{(\Delta v)^2}
\label{equA12}
\end{equation}
where $\Delta v$ is the FWHM of the Gaussian distributions~(\ref{equA1}) and (\ref{equA3}) above, i.e. the width of the grain velocity distribution.

Using equation~(\ref{equA11}), it may be shown that, when $v_{\rm crit} << \Delta v$,

\begin{displaymath}
\langle \sigma v \rangle \approx \sigma \left [\frac {(2\ {\rm ln}\ 2)^3}{\pi }\right ]^{\frac {1}{2}} \frac {v_{\rm crit}^4}{(\Delta v)^3}
\end{displaymath}
\begin{displaymath}
= 0.9209\ \left (\frac {v_{\rm crit}}{\Delta v} \right )^3 \sigma v_{\rm crit}
\end{displaymath}
whereas, when $v_{\rm crit} >> \Delta v$,

\begin{displaymath}
\langle \sigma v \rangle \approx 0.9584\ \sigma \Delta v
\end{displaymath}
In practice, $v_{\rm crit}$ and $\Delta v$ are comparable, and the exact expression~(\ref{equA11}) was used to calculate the coagulation rate coefficient.

In Fig.~\ref{sigmav} is plotted the variation of the rate coefficient for coagulation, $\langle \sigma v \rangle $, with the turbulent width, $\Delta v$, for $a_{\rm g} = 0.05$ $\mu $m. Also shown in the Figure are fractions of the adiabatic sound speed, $c_{\rm s}$ (evaluated at $T = 10$~K). The linear dependence of $\langle \sigma v \rangle $ on $\Delta v$, in the limit $\Delta v << v_{\rm crit}$, is evident in this Figure. The rate coefficient has a maximum near $\Delta v = c_{\rm s}/4$, for the adopted value of $a_{\rm g}$ and assuming that $v_{\rm crit}(a_{\rm g}) = 0.4\ a_{\rm g}^{-5/6}$ (Poppe \& Blum 1997; broken curve in Fig.~\ref{sigmav}); turbulent widths much in excess of or much below this value result in negligible coagulation.

\section{The initial value of $a_{\rm g}$}

Here we demonstrate that the value of $a_{\rm g}$ required to reproduce a given value of $n_{\rm g}\pi a_{\rm g}^2$ depends only on the adopted size distribution and its limits, and not on the density of the grain material nor the dust:gas mass ratio. The grain size distribution is taken from Mathis et al. (1977):

\begin{displaymath}
{\rm d}n_{\rm g} \propto a_{\rm g}^{-3.5} {\rm d}a_{\rm g}
\end{displaymath}
 Let $a_{\rm M}$ be the upper limit to the distribution and $a_{\rm m}$ be the lower limit. Then, following Le Bourlot et al. (1995), the mean grain opacity is given by

\begin{equation}
\langle n_{\rm g} \sigma _{\rm g} \rangle = - \frac {3}{4} \eta n_{\rm H} \frac {a_{\rm M}^{-0.5} - a_{\rm m}^{-0.5}} {a_{\rm M}^{0.5} -  
a_{\rm m}^{0.5}}
\label{equB1}
\end{equation}
where

\begin{equation}
\eta = \frac {1.4 m_{\rm H} G} {\rho_{\rm g}}
\label{equB2}
\end{equation}
and where $\rho_{\rm g}$ is the density of the grain material and $G$ is the dust:gas mass ratio. If, on the other hand, the grains are assumed to have a unique radius, $a_{\rm g}$, then

\begin{equation}
n_{\rm g} \pi a_{\rm g}^2 = \frac {3}{4} \frac {M_{\rm g}} {a_{\rm g} \rho _{\rm g}} 
\label{equB3}
\end{equation}
where $M_{\rm g} = \eta n_{\rm H}\rho_{\rm g}$ is the mass of grains per unit volume of gas. Substituting for $\eta $ from equation~(\ref{equB2}), (\ref{equB3}) becomes

\begin{equation}
n_{\rm g} \pi a_{\rm g}^2 = \frac {3}{4} \frac {\eta n_{\rm H}} {a_{\rm g}} 
\label{equB4}
\end{equation}
Equating equations~(\ref{equB1}) and (\ref{equB4}), we obtain

\begin{equation}
a_{\rm g} = - \frac {a_{\rm M}^{0.5} - a_{\rm m}^{0.5}} {a_{\rm M}^{-0.5} -  
a_{\rm m}^{-0.5}}
\label{equB5}
\end{equation}
which is independent of $\rho _{\rm g}$ and $G$.

\begin{acknowledgements}

GdesF and DRF gratefully acknowledge support from the `Alliance' programme, in 2004 and 2005. We are thankful to Ewine van Dishoeck and Helen Fraser for communicating results of Oberg et al. (2004), in advance of their publication, and to an anonymous referee for a thoughtful and detailed report.

\end{acknowledgements}


\begin{thebibliography}{}
\bibitem[1996]{aikawa96} Aikawa, Y., Miyama, S. M., Nakano, T., Umebayashi, T. 1996, ApJ, 467, 684
\bibitem[1997]{aikawa97} Aikawa, Y., Umebayashi, T., Nakano, T., Miyama, S. M. 1997, ApJ, 486, L51
\bibitem[2001]{aikawa01} Aikawa, Y., Ohashi, N., Inutsuka, S., Herbst, E., Takakuwa, S. 2001, ApJ, 552, 639
\bibitem[2003]{aikawa03} Aikawa, Y., Ohashi, N., Herbst, E. 2003, ApJ, 593, 906
\bibitem[2004]{aikawa04} Aikawa, Y., Herbst, E., Roberts, H., Caselli, P. 2004, astro-ph 0410582
\bibitem[1989]{anders89} Anders, E., Grevesse, N. 1989, Geochim. Cosmochim. Acta, 53, 197
\bibitem[2002]{bacmann02} Bacmann, A., Lefloch, B., Ceccarelli, C., Castets, A., Steinacker, J., Loinard, L. 2002, A\&A, 389, L6
\bibitem[2004]{belloche04} Belloche, A., Andr\'{e}, P. 2004, A\&A, 419, 35
\bibitem[1997]{bergin97} Bergin, E. A., Langer, W. D. 1997, ApJ, 486, 316
\bibitem[2002]{bergin02} Bergin, E.A., Alves, J., Huard, T., Lada, C.J. 2002, ApJ, 570, L101
\bibitem[2003]{bianchi03} Bianchi, S., Goncalves, J., Albrecht, M.
et al. 2003, A\&A, 399, L43
\bibitem[2003]{caselli03} Caselli, P., van der Tak, F. F. S.,
Ceccarelli, C., Bacmann, A. 2003, A\&A, 403, L37 
\bibitem[1993]{chokshi93} Chokshi, A., Tielens, A. G. G. M., Hollenbach, D. 1993, ApJ, 407, 806
\bibitem[2000]{ciolek00} Ciolek, G. E., Basu, S. 2000, ApJ, 529, 925
\bibitem[1987]{draine87} Draine, B. T., Sutin, B. 1987, ApJ, 320, 803
\bibitem[2003a]{flower03a} Flower, D. R., Pineau des For\^{e}ts, G. 2003a, MNRAS, 341, 1272
\bibitem[2003b]{flower03b} Flower, D. R., Pineau des For\^{e}ts, G. 2003b, MNRAS, 343, 390
\bibitem[2004]{flower04} Flower, D. R., Pineau des For\^{e}ts, G., Walmsley, C. M. 2004, A\&A, 427, 887
\bibitem[2004]{geppert04} Geppert, W.D., Thomas, R., Semaniak, J., et al. 2004, ApJ 609, 459
\bibitem[2000]{gibb00} Gibb, E. L., Whittet, D. C. B., Schutte, W. A., Boogert, A. C. A., Chiar, J. E., Ehrenfreund, P., Gerakines, P. A., Keane, J. V., Tielens, A. G. G. M., van Dishoeck, E. F., Kerkhof, O. 2000, ApJ, 536, 347
\bibitem[1985]{hasegawa93} Hasegawa, T. I., Herbst, E. 1993, MNRAS, 261, 83
\bibitem[2003]{kramer03} Kramer, C., Richer, J., Mookerjea, B., Alves, J., Lada C. 2003, A\&A, 399, 1073
\bibitem[1995]{lebourlot95} Le Bourlot, J., Pineau des For\^{e}ts, G., Roueff, E., Flower, D. R. 1995, A\&A, 302, 870
\bibitem[2004]{lee04} Lee, J.-E., Bergin, E.A., Evans, N.J. 2004, ApJ, 617, 360
\bibitem[1985]{leger85} L\'{e}ger, A., Jura, M., Omont, A. 1985, A\&A, 144, 147
\bibitem[1956]{margenau56} Margenau, H., Murphy, G. M. 1956, The Mathematics of Physics and Chemistry (Van Nostrand: Princeton, N.J.)
\bibitem[1977]{mathis77} Mathis, J. S., Rumpl, W., Nordsieck, K. H. 1977,
ApJ, 217, 425
\bibitem[1980]{nakano80} Nakano, T., Umebayashi, T. 1980, PASJ, 32, 613
\bibitem[2004]{oberg04} Oberg, K. I., van Broekhuizen, F., Fraser, H. J., Bisschop, S. E. van Dishoeck, E. F., Schlemmer, S. 2005 ApJ, 621, L33
\bibitem[1993]{ossenkopf93} Ossenkopf, V. 1993, A\&A, 280, 617
\bibitem[1993]{ossenkopf94} Ossenkopf, V., Henning, Th. 1994, A\&A, 291, 943
\bibitem[1961]{osterbrock61} Osterbrock, D. E. 1961, ApJ, 134, 270
\bibitem[1990]{pineau90} Pineau des For\^{e}ts, G., Roueff, E., Flower, D. R. 1990, MNRAS, 244, 668
\bibitem[1997]{poppe97} Poppe, T., Blum, J. 1997, Adv. Sp. Res., 20, 1595
\bibitem[2003]{roberts03} Roberts, H., Herbst, E., Millar, T.J.
2003, ApJ, 591, L41
\bibitem[1996]{savage96} Savage, B. D., Sembach, K. R. 1996, ARA\&A, 34, 279
\bibitem[1987]{shu87} Shu, F. H., Adams, F. C., Lizano, S. 1987, ARA\&A, 25, 23
\bibitem[2001]{sofia01} Sofia, U. J., Meyer, D. M. 2001, ApJ, 554, L221
\bibitem[1978]{spitzer78} Spitzer, L. 1978, Physical Processes in the Interstellar Medium (Wiley: New York)
\bibitem[2001]{suttner01} Suttner, G., Yorke, H. W. 2001, ApJ, 551, 461 
\bibitem[2004]{tafalla04} Tafalla, M., Myers, P. C., Caselli, P., Walmsley, C. M. 2004, A\&A, 416, 191
\bibitem[2004]{vastel04} Vastel, C., Phillips, T. G., Yoshida, H. 2004, ApJ, 606, L127
\bibitem[2004]{wakelam04} Wakelam, V., Caselli, P., Ceccarelli, C., Herbst, E., Castets, A. 2004, A\&A, 422, 159 
\bibitem[2004]{walmsley04} Walmsley, C. M., Flower, D. R., Pineau des For\^{e}ts, G. 2004, A\&A, 418, 1035
\end{thebibliography}
\end{document}